\documentstyle[12pt]{article}
\begin{document}
\pagestyle{empty} \large \noindent
G\"{o}teborg ITP 92-17\\
April 1992\\
\begin{center} \LARGE \bf \vspace*{40mm}
Ashtekar's variables for arbitrary gauge group\\
\vspace*{20mm} \Large \bf
Peter Peld\'{a}n\\
\vspace*{5mm} \large
Institute of Theoretical Physics\\
S-412 96  G\"{o}teborg, Sweden\\
\vspace*{40mm} \Large \bf
Abstract\\
\end{center} \normalsize
A generally covariant gauge theory for an arbitrary gauge group with dimension
$\geq 3$,
 that reduces to Ashtekar's canonical formulation of gravity for SO(3,C), is
presented.
 The canonical form of the theory is shown to contain only first class
 constraints.

\newpage \pagestyle{plain}

When Ashtekar \cite{Ash} managed to reformulate Einstein gravity on Yang-Mills
phase
space, it rekindled the old dream of finding a unified theory of gravity and
Yang-Mills
theory. However, it soon became clear that this Ashtekar formulation relied
heavily on the use of the gauge group SO(3) (or locally isomorphic ones), and
the
simple structure-constant identity that exists for these groups. Without this
identity the constraint algebra fails to close, and the theory is not
diffeomorphism invariant, and contains second class constraints. In an attempt
to
find an Ashtekar formulation for an arbitrary gauge group, there are no
problems with
 the generator of gauge transformations (Gauss' law), or the generator of
spatial
diffeomorphisms (the vector constraint). They
form a system of first class constraints by themselves, for arbitrary gauge
group. The
difficult part is the generator of diffeomorphisms off the spatial
hyper-surface (the
Hamiltonian constraint). This constraint is constructed with the help of the
structure
constants, and in the Poisson bracket between two Hamiltonian constraints, the
identity, mentioned  above, is needed to give a weakly vanishing result. So,
the
strategy to construct the general theory
is :Write down a Hamiltonian constraint without the use of the
structure constants, such that, when choosing the gauge group SO(3), the
constraint
reduces to the ordinary Ashtekar constraint. The hope is then that the
construction
works for an arbtrary gauge group, since one does not use any particular
feature of a
special gauge group any more. To do this in practice, first define a scalar
with the
help of the four fundamental scalar densities :\normalsize
$\epsilon_{abc}f_{ijk}\Pi^a_i\Pi^b_jB^c_k$,
$\epsilon_{abc}f_{ijk}\Pi^a_i\Pi^b_j\Pi^c_k
$, $\epsilon_{abc}f_{ijk}\Pi^a_iB^b_jB^c_k$,
$\epsilon_{abc}f_{ijk}B^a_iB^b_jB^c_k$.
Then, multiply the ordinary Ashtekar Hamiltonian constraint with this scalar,
and,
finally, use the structure-constant identity to eliminate all structure
constants.
This new Hamiltonian will then in general give a closed constraint algebra for
an
arbitrary gauge group.

 In this letter, I will show how to obtain
this Ashtekar theory for an arbitrary gauge group, through a Legendre transform
from a
 pure connection Lagrangian of the form discovered by Capovilla, Jacobson
and Dell\cite{CDJ1}. The resulting canonical theory will correspond to
multiplying the
Hamiltonian constraint by the determinant of the "magnetic" field, in the
strategy
above.

  In order to find the Ashtekar-theory for a general gauge group, I will start
with the
generally covariant and gauge invariant CDJ-action \cite{CDJ1}.
 \begin{equation}
S  =  \frac{1}{8} \int d^4x\;  \eta (Tr \Omega^2 + a (Tr \Omega)^2) \label{1}
\end{equation}  \\
where $\Omega^{ij}:= \epsilon^{\alpha \beta \gamma \delta} F_{\alpha \beta}^{i}
F_
{\gamma \delta}^j$ and $ F_{\alpha \beta}^i = \partial_{[ \alpha} A_{\beta]}^i
+
f_{ijk} A_{\alpha}^j A_{\beta}^k$ and $\eta$ is a scalar density of weight
minus one.
 The trace is taken with the invariant bilinear Killing-form of the
Lie-algebra. (For
some Lie-algebras, like so(1,3), so(4) and so(2,2), there exist two different
"traces"
that could be used in the action (\ref{1}), meaning that for these groups,
there
exist an even more general Lagrangian, quartic in the field-strength
\cite{PP2}).
 A 3+1 canonical decomposition of this
action, with $a = - \frac{1}{2}$ and the gauge group SO(3,C), is known to give
Ashtekar's Hamiltonian for pure gravity without cosmological constant. For
other
values of $a$ but keeping the gauge group SO(3,C), the action still describes a
theory
that has an interpretation in terms of Riemannian geometry
\cite{Capovilla,Bp1}.
 This pure connection
formulation of general relativity has also been studied with cosmological
constant and
matter couplings in three and four dimensions \cite{CDJ1,PP3,PP}.

However, no one has yet given the canonical form for arbitrary gauge group.
 Since this action is
invariant under diffeomorphisms and gauge transformations, it should be quite
clear
that a canonical decomposition of it should give a set of first class
constraints
generating these symmetries. And since one knows that with the gauge group
SO(3,C) it
 gives Ashtekar's variables, the general theory must be what one would call
"Ashtekar's variables for arbitrary gauge group". However, there are at least
two
things that could ruin this construction. The first is that it could be
"impossible"
to perform the Legendre transform for a general gauge group. (This is not so
strange
since previous work on this action has relied heavily on the fact
that the gauge group is
three-dimensional, and that the structure constants satiesfies the simple SO(3)
identity.) The other thing that could have ruined the beauty of the Hamiltonian
formulation, is that complicated second class constraints would have appeared.
However,
none of these are the case, and as I soon will show, the only thing that
happens for an
arbitrary gauge group, is that the Hamiltonian constraint splits up into three
pieces.

First I define the momenta conjugated to $A_a^i$.
\begin{equation}
\Pi_{i}^a := \frac{ \partial {\cal L}}{\partial \dot{A_{a}^i}}=\eta
(\Omega^{i}_{j} + a
Tr\Omega \delta^{i}_{j})B^{aj}  \label{2} \end{equation}\\
where $B^{aj}:=\epsilon^{abc} F^j_{bc}$ is the "magnetic field". a,b,c denote
spatial
indices and i,j,k denote gauge indices. Now, it is rather straightforward to
perform
the Legendre transform, provided the "magnetic metric"  $b^{ab}:=B^{ai}B^b_i$
is
invertible. It is here that one must require the dimension of the gauge group
to be
$\geq 3$ in order to have a non-degenerate "magnetic metric". Defining the
inverse of
the "magnetic metric" : $b_{ab}:= \frac{1}{2 det(b^{cd})}
\epsilon_{aef}\epsilon_{bgh}b^{eg}b^{fh}$, and performing the Legendre
transform gives

 \begin{equation}
{\cal H}_{tot}=N {\cal H} + N^a {\cal H}_a + \Lambda^i {\cal G}_i \label{3}\\
\end{equation}
where\\
\normalsize \begin{eqnarray}
{\cal H}&=& \frac{\sqrt{det(b^{ab})}}{4}( 2 \Pi^{ai} \Pi^b_ib_{ab} -
(\Pi^{ai}B^b_i)b_{bc}(\Pi^{cj}B^d_j)b_{da}-\frac{a}{1+3a}
(\Pi^{ai}B^b_ib_{ab})^2)\approx0\nonumber\\
{\cal H}_a&=&\frac{1}{2} \epsilon_{abc} \Pi^{bi} B^c_i\approx0\nonumber\\
{\cal G}_i&=&{\cal D}_a \Pi^a_i \approx0\nonumber\\
N&:=&\frac{1}{2\eta\sqrt{det(b^{ab})}}\nonumber\\
\Lambda^i&:=&-A_0^i\nonumber\end{eqnarray}
${\cal H}$ is usually called the Hamiltonian constraint, ${\cal H}_a$ the
vector
constraint and ${\cal G}_i$ Gauss' law. The value $a=-\frac{1}{3}$ must be
handled
separetely. For that case, the Hamiltonian constraint splits up in two separate
pieces which will become second class constraints.
 Notice that the form of Gauss' law and the
vector
constraint are independent of gauge group while the Hamiltonian constraint
looks a bit
more complicated for a general gauge group compared with the ordinary Ashtekar
form
for SO(3):  $
{\cal H}_{Ash}=\frac{i}{4}\epsilon_{abc}f_{ijk}\Pi^a_i\Pi^b_jB^c_k$.
It is  however easy to check that with $a=-\frac{1}{2}$ and the gauge group
SO(3), using the identity
$f^{ijk}f_{lmn}=\delta^{[i}_l\delta^{j]}_m\delta^k_n+\delta^{[i}_n\delta^{j]}_l
\delta^k_m + \delta^{[i}_m\delta^{j]}_n\delta^k_l$, valid for SO(3), ${\cal H}$
can be
rewritten as
 ${\cal H}=i {\cal H}_{Ash}$.

 Now, for an arbitrary gauge group, one must still check that the constraints
 form a first
class set. And, as mentioned earlier, there are no problems with Gauss' law and
the
vector constraint. We know that they generate gauge transformations and spatial
diffeomorphisms, and all constraints are gauge covariant and diffeomorphism
covariant,
which means that all Poisson brackets including these constraints are weakly
vanishing. So, the only non-trivial calculation is the Poisson bracket between
two
Hamiltonian constraints. A straightforward calculation gives
 \begin{equation} \{{\cal H}[N],{\cal H}[M]\}={\cal
H}_a[q^{ab}(N\partial_bM-M\partial_bN)]\label{5} \end{equation}
where  \begin{eqnarray}{\cal H}[N]=\int d^3x\;{\cal H}N \nonumber
 \end{eqnarray}
and \normalsize
 \begin{eqnarray}q^{ab}&=& 2 \Pi^{ai}\Pi^b_i - 3
(\Pi^{ai}B^c_i)b_{cd}(B^d_j\Pi^{bj})+ \frac{1\!+\!2a}{1\!+\!3a}
(b_{cd}\Pi^{ci}B^d_i)
\frac{1}{2} (\Pi^{(aj}B^{b)}_j)-\nonumber\\&& -
\frac{(1\!+\!a)(\frac{1}{2}\!+\!a)}{(1+3a)^2}
(b_{cd}\Pi^{ci}B^d_i)^2 b^{ab} + 3(b_{ef}(\Pi^{ei}B^c_i)b_{cd}(B^d_j\Pi^{fj})\!
-
\!b_{ef} \Pi^{ei}\Pi^f_i) b^{ab} \nonumber\end{eqnarray}
on the constraint surface. And, according to Hojman, Kucha\v{r} and Teitelboim
\cite{HKT}
the object $q^{ab}$ in the Poisson bracket above is to be interpreted as
the spatial metric on the
hyper-surface.

 From now on I will put $a=-\frac{1}{2}$ which is the value that for SO(3,C)
gives
ordinary gravity. This means that the spatial metric is
\normalsize \begin{eqnarray} q^{ab}&=& 2 \Pi^{ai}\Pi^b_i - 3
(\Pi^{ai}B^c_i)b_{cd}(B^d_j\Pi^{bj})+\\&&+
3(b_{ef}(\Pi^{ei}B^c_i)b_{cd}(B^d_j\Pi^{fj})
 -
b_{ef} \Pi^{ei}\Pi^f_i) \;b^{ab} \nonumber\end{eqnarray}
a form that makes it very hard to ensure positive definiteness of the metric,
for a
general gauge group.

During the calculation of the Poisson bracket (\ref{5}), it becomes clear that
the
only parts that could give a non-closure of the constraint algebra, come from
the
first term in the Hamiltonian constraint. That means that there exists another
Hamiltonian constraint, quadratic in momenta, that gives a closed constraint
algebra,
 namely
\normalsize \begin{eqnarray}{\cal H}^{Alt.}&=&\frac{\sqrt{det(b^{ab})}}{4}
((\Pi^{ai}B^b_i)b_{bc}
(\Pi^{cj}B^d_j)b_{da} -
(\Pi^{ai}B^b_ib_{ab})^2)=\\&=&-\frac{1}{4\sqrt{det(b^{ab})}}
\epsilon_{abc}\epsilon_{def}(\Pi^{ai}B^d_i)(\Pi^{bj}B^e_j)b^{cf}\nonumber\end{eqnarray}
 This Hamiltonian constraint seems more tractable than the original one in
 (\ref{3}),
at first sight. However, it has two remarkable features. First, doing the
Legendre
transform backwards for this Hamiltonian gives not a manifestly covariant pure
connection action, despite the fact that one has a closed constraint algebra.
(The same situation appears for
gravity coupled to a massive spinor in the ordinary Ashtekar formulation
\cite{PP}).
Second, the theory only cares about the part of $\Pi^{ai}$ that is
non-orthogonal to
$B^{ai}$ in its internal gauge indices. The orthogonal part of $\Pi^{ai}$ do
not have
any effect on $A^i_a$ at all. Which Hamiltonian constraint is then the best
choice for
a generalization of Ashtekar's variables? Both reduces to the Ashtekar
Hamiltonian
constraint for SO(3), ${\cal H}^{Alt}$ has a simpler form, but ${\cal H}$
corresponds
to a manifestly covariant Lagrangian. My opinion is that the pure connection
Lagrangian (\ref{1}) really has some fundamental role, and therefore one should
choose
${\cal H}$. And, besides that, it is something strange with a theory which has
a lot
of "extra" fields (the "orthogonal" part of  $\Pi^{ai}$)  which have no effect
on the
equations of motion.

If one is only looking for a theory that for gauge group SO(3) reduces to the
Ashtekar formulation, and does not mind whether, for instance, the Hamiltonian
is
quadratic
in momenta or not, then there exist several different Hamiltonian constraints.
They can
be found in three different ways. 1.Use the strategy outlined in the beginning
of this
letter. 2. Write down the general CDJ-type Lagrangian with arbitrarily high
orders in
the field-strength, and perform the Legendre transform. 3. Define the
Hamiltonian
constraint by contracting the following scalar densities with epsilon-tensor
densities:
$\Pi^{ai} B^b_i$, $B^{ai} B^b_i$ and $\Pi^{ai} \Pi^b_i$. Here is an
example of the third way:
\normalsize ${\cal
H}=\epsilon_{abc} \epsilon_{def}(\Pi^{ai} \Pi^d_i)(\Pi^{bj}\Pi^e_j)(\Pi^{ck}
B^f_k)$.
 This is an interesting Hamiltonian constraint, which has the feature that the
spatial metric will be of the familiar form $q^{ab} \sim \Pi^{ai} \Pi^b_i$.
Using
the third way, one must carefully check the constraint algebra,
it will not always be closed.

Now, the existence of a theory for a general gauge group, which reduces to the
theory of
 pure
Einstein gravity for a specific choice of the gauge group, makes it tempting to
speculate about a unified description of gravity and Yang-Mills theory. The
question is
then: What kind of gauge group should be used? Could the naive guess of
$SO(3)\times G$
give gravity coupled to a Yang-Mills field with gauge group $G$, or is there
need for
a more sophisticated construction that in some way could be reduced to the
desired
result (spontaneos symmetry breakdown?)? The first thing one may notice is
 that with gauge
group $SO(3)\times G$ the Hamiltonian in (\ref{3}) can never give
 the "ordinary" coupling of
gravity-Yang-Mills, given by
Ashtekar, Romano and Tate \cite{ART} (That is because, the ordinary coupling is
 non-homogeneous in the momenta, while ${\cal H}$ in (\ref{3}) is just
quadratic.).
Perhaps, some of the other generalized Hamiltonians, mentioned above, have a
better
chance. However,  what is required of the coupling is really
just that it reduces to the ordinary Yang-Mills equations for flat space-times
in the
weak field limit. But, even that seems to fail. Trying the  gauge group
 $SO(3)\times  U(1)$, it is easy to verify that Maxwell's equations do not
appear
in the weak field limit. So, if there will be no miraculous improvements for
some
special Yang-Mills gauge group, this naive construction will fail, and one
must really think of something more clever.

Also the reality condition of the Ashtekar formulation seems tough to handle in
this
direct-product approach. In general, the reality condition will have to be
 matter field dependent, in order to get a real metric.

However, an optimistic speculation regarding the reality conditions, is that it
could
be possible to find a gauge group in which the "gravity part" would give a
positive
definite spatial metric without any need of introducing complex fields.

Looking ahead a bit, one could start thinking of using the Rovelli-Smolin
loop-representation quantization scheme \cite{RS} for this generalized theory.
At
first sight, it is two obvious things that change: The "spinor identity" will
become
more complicated, and the definition of the hamiltonian constraint in terms of
the
T-variables and T-operators changes. But otherwise it does not seem to be an
impossible task to redo everything for a general gauge group.\\

I would like to thank Ingemar Bengtsson for numerous discussions.

\end{document}